\documentclass[11pt]{article}
\usepackage{moriond,epsfig}

\def\f{\phi}

\def\j{\psi}

\def\n{\nu}

\def\F{\Phi}

\def\tM{\widetilde{M}}

\def\beq{\begin{equation}}
\def\eeq{\end{equation}}
\def\bea{\begin{eqnarray}}
\def\eea{\end{eqnarray}}

\def\pl#1#2#3{Phys.~Lett.~{\bf B {#1}} ({#2}) #3}
\def\np#1#2#3{Nucl.~Phys.~{\bf B {#1}} ({#2}) #3}

\def\pr#1#2#3{Phys.~Rev.~{\bf D {#1}} ({#2}) #3}

\begin{document}

\vspace*{4cm}
\title{Proton Decay in a 6D $SO(10)$ model}

\author{L. COVI }

\address{Theory Division, CERN Department of Physics, Geneva,
    Switzerland}

\maketitle\abstracts{
We present a study of proton decay in a supersymmetric $ SO(10)$ 
gauge theory in
six dimensions compactified on an orbifold. The dimension 5 proton
decay operators are absent, but the dimension 6
are enhanced due to the presence of KK towers.  
We resum the KK modes up to the cut-off of the theory and find
the rate for the dominant mode $p \rightarrow \pi^0 e^{+} $. 
We explore also the flavour dependence, due to the different
localization of states in the extra dimensions and find that it
is possible to distinguish the model from the usual 4D $ SU(5)/SO(10)$ 
models.
}

Proton decay is still the main signature of Grand Unified Theories as
it was realized more than 30 years ago \cite{old-proton}.
Unfortunately for GUT model-builders, it is still escaping observation...,
but it is perhaps not surprising since the lifetime scales as the 
fourth power of the GUT scale and the experimental limits can be improved 
only at the cost of huge detectors. 
At present the most stringent experimental bounds come from the
SuperKamiokande collaboration, that at this conference announced
some new limits, e.g. 
$\tau (p\rightarrow e^+\pi^0)\geq 6.9\times 10^{33} $ years
and $\tau (p\rightarrow K^+\bar\nu)\geq 1.6\times 10^{33} $ 
years\cite{SK-moriond}. 
Such values exclude simple
non-supersymmetric $SU(5)$ models~\footnote{Actually such models do not
fare well with the unification of the gauge couplings either, 
as we all know, so probably excluding them is superfluous...} 
and start to reduce the parameter space also of the supersymmetric 
version, depending on the flavour
structure assumed~(see e.g.\cite{dim5}).

In this talk I will describe the results for proton decay in a 
particular orbifold model computed in~\cite{bccw04}. 
I will try to argue both that the signal 
could really be around the corner in this case and also that 
{\it if } the signal is seen, we could perhaps have a chance to 
distinguish if the GUT model is of the 4 dimensional type or 
presents some extra-dimensional features.

\section{$SO(10)$ model in 6D}

Our starting point is an $ SO(10) $ gauge theory in 6D with $N=1$
supersymmetry compatified on the orbifold $T^2/(Z^I_2\times
Z^{PS}_2\times Z^{GG}_2)$ \cite{abc01,hnx02}. The theory has four
fixed points, $O_I$, $O_{PS}$, $O_{GG}$ and $O_{fl}$, located at the
four corners of a `pillow' corresponding to the two compact dimensions.  
At $O_I$ only supersymmetry is broken whereas at the
other fixed points, $O_{PS}$, $O_{GG}$ and $O_{fl}$, also the gauge
group $ SO(10) $ is broken to its three GUT subgroups G$_{PS}$=
$SU(4)\times SU(2) \times SU(2) $, 
G$_{GG} = SU(5) \times U(1)_X $ 
and flipped $SU(5)$, 
G$_{fl} = SU(5)'\times U(1)' $,
respectively. The intersection of all these GUT groups yields the
standard model group with an additional $ U(1) $ factor, G$_{SM'}=
 SU(3) \times SU(2) \times U(1)_Y \times U(1)_{Y'}$, 
as unbroken gauge symmetry below the compactification
scale.  The field content of the theory is strongly constrained by
imposing the cancellation of irreducible bulk and brane anomalies
\cite{abc03}.  We study the model proposed in \cite{abc03b},
containing 3 {\bf 16}-plets $\j_i$, $i=1\ldots 3$, as brane fields and
6 {\bf 10}-plets, $H_1,\ldots, H_6$, and 4 {\bf 16}-plets, $\F, \F^c,
\f, \f^c$, as bulk hypermultiplets.  Vacuum expectation values of $\F$
and $\F^c$ break the surviving $ U(1)_{B-L}$.  The electroweak
gauge group is broken by expectation values of the anti-doublet and
doublet $H_{u/d} $ contained in $H_1$ and $H_2$.

\begin{figure}
\begin{center}
\psfig{figure=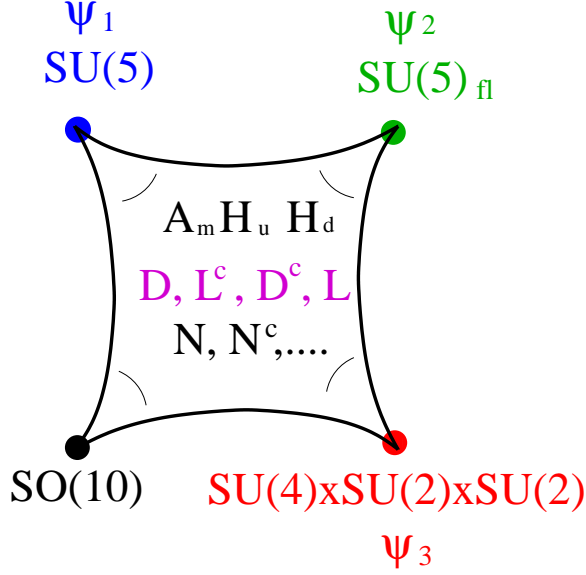, height=3in}
\caption{
The orbifold structure in the extra dimensions and the
localization of the fields.
\label{fig:pillow}}
\end{center}
\end{figure}

We choose the parities of $H_5$, $H_6$ and $\f,\f^c $ such that their
zero modes 
\begin{equation}
  L = \left( \begin{array}{l} 
      \n_4 \\ e_4
    \end{array} \right)\;, \quad  L^c = \left( \begin{array}{l} 
      \n^c_4 \\ e^c_4 
    \end{array}\right)\;, \quad D^c = d^c_4\;, \quad D = d_4\;
\end{equation}
act as a (vectorial) fourth generation of d-quarks
and leptons and mix with the three generations of brane fields,
located on the three branes where $ SO(10)$ is broken to its three 
GUT subgroups.  
This leads to a characteristic pattern of mass matrices of the
lopsided type as described in \cite{abc03b}. In particular
the hierarchy and mixing angles for the quark sector can be
accounted for and GUT relations do not hold for all the
generations due to the presence of split multiplets. 
See~\cite{abc03b} for the details
and~\cite{bccw04} for the explicit form of the mixing matrices.

\section{Short review of 4D proton decay in supersymmetry}

Proton decay arises from effective 4-fermion operators 
joining three quarks and a lepton, which are of dimension 6. 
It can therefore be a bit puzzling to hear about ``dimension 4'' 
or ``dimension 5'' operators, as happens in supersymmetric models. 
So I will stop a little and review the terminology before discussing 
the dominant contribution in our case.

In supersymmetric models, there are contributions to proton 
decay from superpotential terms, either renormalizable or obtained 
integrating out heavy states, from kinetic terms and also from 
supersymmetry breaking terms. 
The first type of contributions, are usually classified
according to the dimension of the superpotential
terms that break the baryon or lepton number.
In general we have 
\begin{itemize}

\item{dimension four operators:
\beq
W = \lambda L L E^c + \lambda' L Q D^c + \lambda^{''} U^c D^c D^c ;
\eeq
they are renormalizable and give very rapid proton decay 
via an intermediate
scalar squark (therefore the effective 4-fermion operator is
just suppressed by ${\lambda'\lambda^{''} \over m^2_{susy}} $,
where $m^2_{susy} $ is the typical squark mass) and have to be 
excluded by a discrete symmetry, usually R-parity.
}

\item{dimension five operators:
\beq
W = {1\over M_{H_C}} 
\left[
{1\over 2} Y_{qq} Y_{ql} Q Q Q L 
 + Y_{ue} Y_{ud} E^c U^c U^c D^c ;
\right]
\eeq
they arise e.g. from integrating out the heavy colored Higgs triplets with 
mass $M_{H_C} $ and allow the decay via a loop of scalar superpartners. 
They produce effective 4-fermion operators that scale as 
${1\over M_{H_C} m_{susy}} $, are color antisymmetric and therefore
must be also flavour non-diagonal, so that the dominant channel results in
$p \rightarrow K^{+} \bar\nu $.
Such operators give the dominant contribution in the simple
supersymmetric SU(5) case, and they can give proton decay even above
the present limit~\cite{dim5}.
}

\item{``real'' dimension 6 operators, arising from the fermion 
kinetic terms and mediated by the gauge multiplet; they are therefore 
not of the chiral type and cannot be written as superpotential terms.
They do not involve sparticles and are therefore independent of $m_{susy}$,
apart for the weak dependence coming from determining the GUT scale by RGEs.
As an example, in $ SU(5)$ we have the exchange of the $ {\cal X} $
leptoquark gauge bosons with masses $M_{\cal X}$. 
The effective vertex is give by the Fermi-type coupling:
\begin{equation}
  {\cal L}_{eff} = - \frac{g_5^2 }{ 2 M_{\cal X}^2}\;
  \epsilon_{\alpha\beta\gamma}\;\overline{u^c}_{\!\!\alpha, i}\,
  \gamma^\mu\, Q_{\beta,i}\; \left[ \overline{e^c}_{\!\!j}\,
    \gamma_\mu\, Q_{\gamma,j}\, - \overline{d^c}_{\!\!\gamma,k}\,
    \gamma_\mu\, L_k \right] + \mbox{h.c}. \; ,
  \label{eff4D}
\end{equation}
where $i,j$ and $k$ count the generations.
With Fierz reordering, one can write the operators in the usual
form as
\begin{equation}
  {\cal L}_{eff} = - \frac{g_5^2 }{ M_{\cal X}^2}\;
  \epsilon_{\alpha\beta\gamma} \left[ \overline{e^c}_{\!j}
    \overline{u^c}_{\!\alpha,i}\, Q_{\beta,i}\, Q_{\gamma,j} -
    \overline{d^c}_{\!\alpha,k} \overline{u^c}_{\!\beta,i}\,
    Q_{\gamma,i}\, L_k \right] + \mbox{h.c.}  \, .
  \label{effWeyl-RRLL}
\end{equation}
Note that these operators scale as $ M_{\cal X}^{-2}$ and are 
proportional to a gauge coupling and not a Yukawa;
still some flavour dependence arises from the quark and
lepton mixing matrices.
Assuming quarks and leptons to be embedded into multiplets
according to their hierarchy, the dominant decay channel is the one
involving only first generation fermions, i.e.
$ p \rightarrow \pi^0 e^{+} $.
}

\item{dimension 6 operators coming from supersymmetry breaking,
e.g. the one mediated by intermediate Higgs scalars mixing
via the soft SUSY breaking mass terms;
they are usually much more suppressed compared to the
previous ones and are usually neglected.
}
\end{itemize}

\section{6D proton decay}

In our 6D orbifold model we have a residual 4D N=1 supersymmetry and
we could have in principle dimension 4 and dimension 5 proton decay.
Luckily in extra-dimensional models, we can exclude them both 
with appropriate choice of R-symmetry, forbidding the $\lambda',\lambda''$
couplings and also the $\mu$ term \cite{abc03b}.
Note also that in extra-dimensional models, the two heavy triplet Higgs 
bosons become massive together with their N=2 superpartners, not directly 
with each other, so the mixing between them is generated only by
supersymmetry breaking. 
So in general the dominant contribution to proton decay in orbifold
models comes from dimension 6 operators \cite{dim5orbifold}.
Moreover in our model, since the first generation of $u$ quarks are
confined to live on the fixed point where $SU(5)$ is unbroken, 
we can use $SU(5)$ language to
describe the operators, even if the bulk symmetry is $SO(10)$~\footnote{
A small effect from the other gauge bosons can arise from brane
derivative operators \cite{bccw04}.}.

\subsection{Effective operator in 6D}

In our orbifold model there is an important difference compared to
the 4D case, we have to take into account the presence of a 
Kaluza-Klein tower of {\cal X} gauge bosons with masses given by
\begin{equation}
  M_{\cal X}^2 (n,m) = \frac{(2n+1)^2}{R_5^2} + \frac{(2m)^2}{R_6^2} 
\end{equation}
for $n,m = 0$ to $\infty$.  The lowest possible mass is
$M_{\cal X}(0,0)=1/R_5$, as given by the $SU(5)$ breaking parity
\cite{abc01,hnx02}.
Note that if we define the 4D gauge coupling as the effective coupling
of the zero modes, the KK modes interact more strongly by a factor
$\sqrt{2} $ due to their bulk normalization.

To obtain the low energy effective operator, we have then to sum over
the Kaluza Klein modes. We can define
\begin{equation}
  \frac{1}{(M_{\cal X}^{eff})^2} = 2 
  \sum_{n,m=0} \frac{1}{ M_{\cal X}^2 (n,m)} 
=  2 \sum_{n,m=0} \frac{ R_5^2}{ (2 n +1)^2 + \frac{R_5^2}{ R_6^2}
    (2 m)^2} \;\; ;
  \label{sumKK}
\end{equation}
taking the limit $R_6/R_5 \rightarrow 0$, we regain the finite 
5D result \cite{hm02},
\begin{eqnarray}
  2 \sum_{n=0}^{\infty} \frac{R_5^2}{(2n+1)^2} = \frac{\pi^2 R_5^2}{4}\;
  . 
  \label{eq:coupling-constant}
\end{eqnarray}

But in 6D the summation shows a logarithmic divergence; since our
theory is non-renormalizable and valid only below the scale $M_*$, 
where the theory becomes strongly coupled and 6D gravity corrections 
are no more negligible, we regulate the sum with the cut-off $M_*$, 
and obtain formally
\begin{eqnarray}
  \frac{1}{(M_{\cal X}^{eff})^2} 
  \simeq \frac{\pi}{ 4}\, R_5 R_6 \left[\, 
\ln \left( M_* R_5 \right) + C\left(\frac{R_5}{R_6} \right) 
+ {\cal O} \left(\frac{1}{R_{5/6} M_*}\right)\;
\right]\; .
\end{eqnarray}

In the case $R_5 = R_6 = 1/M_c $ the expression can be approximated by
\begin{eqnarray}
\frac{1}{(M_{\cal X}^{eff})^2} \simeq 
\frac{\pi}{ 4 M_c^2}\,  \left[\, \ln \left( \frac{M_*}{M_c} \right) + 
2.3 \right] \; ,
\end{eqnarray}
which agrees within 1\% with explicit discrete sum for
$\frac{M_*}{M_c} = 10\dots 50 $.

\section{Flavour structure}

Another important difference in 6D is the non-universal coupling of the
${\cal X}$ gauge bosons.  In fact, due to the parities and the $ SO(10)$ 
breaking pattern, their wavefunctions must vanish on the
fixed points $O_{PS}$ and $O_{fl}$, and
therefore no coupling arises with the charm and
top quark or to the brane states $d^c_{2,3}, l_{2,3}$.  We have in
principle couplings to the bulk states $d^c_4,d_4$ and $l_4, l^c_4$,
but in this case, the charge current interaction always mixes the light 
states with the heavy KK modes and it is therefore irrelevant for the 
low energy proton decay~\cite{hm02}. 
So the kinetic coupling in Eqn.~(\ref{effWeyl-RRLL})
arises only for the $1st$ flavour eigenstate, not for all flavours as in
the usual $4D$ case.

Proton decay involves only the light quark states and the operators
containing the combinations $uud$ and $udd$. Starting in the basis where
the up-quark Yukawa is diagonal, we have to rotate the down-type quarks 
and the leptons from the weak into the mass eigenstates and single out
the contributions for the lightest generation. We have then
\begin{eqnarray}
  d_L  = U^d_L d'_L\;\;, e_L = U^e_L e'_L\;\;,  \nu_L  = U^{\nu}_L
  \nu'_L \;\;,  d_R  = U^d_R d'_R\;\;, e_R = U^e_R e'_R\;,
\end{eqnarray}
where the prime denotes mass eigenstates. Since the up quarks are
diagonal, $U^d_L$ coincides with the CKM-matrix.
We can write the proton decay operators of Eqn.~(\ref{effWeyl-RRLL}) in
mass eigenstates as
\begin{eqnarray}  \label{eff6D-VU}
  {\cal L}_{eff} &=& \frac{g_5^2}{ (M_{\cal X}^{eff})^2}\;
  \epsilon_{\alpha\beta\gamma} \left[ 2\, \overline{e^c}'_{\!\!k}
    \left( U^{e\top}_R \right)_{k1} \overline{u^c}_{\!\!\alpha,1} \,
    d'_{\beta,m} \left( U^d_L \right)_{1m} u_{\gamma,1} \right.
  \\
& &  + \left. \overline{d^c}'_{\!\!\alpha,l} \left( U^{d\top}_R
    \right)_{l1} \overline{u^c}_{\!\beta,1} \left( u_{\gamma,1} \left(
        U^e_L \right)_{1j} e'_j - d'_{\gamma,m} \left( U^d_L
      \right)_{1m} \left(U^{\nu}_L \right)_{1j} \nu'_j \right) \right]
  + \mbox{h.c.}
\end{eqnarray}
Note again that due the orbifold construction  only the first
weak eigenstates couple to the ${\cal X}$ bosons, instead of all of them.
So the proton decay in this 6D model has naturally
different branching ratios compared to a 4D model {\it with the
same mixing matrices}.

\section{Results}

\subsection{Bound on $M_c$}

Considering the dominant channel $p\to e^+ \pi^0$,
a lower bound on the compactification scale can be derived from the
SuperKamiokande limit on the lifetime.  
We have in fact
\begin{equation}
  \Gamma \simeq K_{had}^{\pi^0}\, \frac{\pi^2}{16}{M_*^4\over M_c^4}
  \left( \ln \left( \frac{M_*}{M_c}\right) + 2.3 \right)^2
  \left[\, 4 V_{ud}^4 + \frac{\tM_2^{d\,2}}{\tM_1^{d\,2}+\tM_2^{d\,2}}
    \frac{\tM_2^{e\,2}}{\tM_1^{e\,2}+\tM_2^{e\,2}} \right]\; ,
\end{equation}
where $K_{had}^{\pi^0} = 1.87 \times 10^{-40}$ sec$^{-1}$ contains 
a factor $M_*^{-4} $, the 
hadronic matrix element, kinematical factors, gauge coupling and 
the running of the operator from the high to the proton scale~\cite{bccw04}.
With $M_*=10^{17}\,\mbox{GeV}$ and $\tM^{d,e}_{2}= 0 $, 
the limit $\tau\geq 6.9\times 10^{33}$ yields
$M_c \geq 0.89\times 10^{16}\,\mbox{GeV}$, not far from the 4D GUT
scale.

\subsection{Rates and branching ratios}

\begin{table}[t]
  \centering
  \begin{tabular}{l|rr|r}
    decay channel & \multicolumn{3}{c}{Branching Ratios [\%]} \\
    \cline{2-4}
    & \multicolumn{2}{c|}{6D $SO(10) $} &
    $ SU(5)\times U(1)_F$ \\ 
    & case I & case II &  models A \& B\\ 
    \hline
    $e^+\pi^0$ & 75 & 71 & 54 \\
    $\mu^+\pi^0$ & 4 & 5 & $\leq $\,1 \\
    $\bar\nu\pi^+$ & 19 & 23 & 27 \\
    $e^+ K^0$ & 1 & 1 & $\leq $\,1 \\
    $\mu^+ K^0$ & $\leq $\,1 & $\leq $\,1 & 18 \\
    $\bar\nu K^+$ & $\leq $\,1 & $\leq $\,1 & $\leq $\,1 \\
    $e^+\eta$ & $\leq $\,1 & $\leq $\,1 & $\leq $\,1 \\
    $\mu^+\eta$ & $\leq $\,1 & $\leq $\,1 & $\leq $\,1 \\
  \end{tabular}
\caption{\label{tb:result}
Resulting branching ratios and comparison with $ SU(5)\times U(1)_F$.
See~\protect\cite{bccw04} for the details of the computation.
}
\end{table}

We calculate the branching ratios for dimension 6 proton decay in
our model and in some lopsided 4D models with a similar flavour 
structure discussed in~\cite{Altarelli02}. As expected,
we find sizable differences in many channels, most noticeably in
$p\rightarrow \mu^+ K^{0} $ due to the absence of direct coupling 
of the second generation weak eigenstates to the
${\cal X}$ gauge bosons.
Even changing the unknown high energy parameters does not modify
the picture: if we vary the heavy masses
$\widetilde{M}_j/\widetilde{M}=0.1,\ldots,1$ and
$\tilde\mu_3^{d,e}/\mu_3=1,\ldots,5$, we still find 
$BR(p\to \mu^+ K^0) \leq 5\,\% $.

\section{Conclusions} 

We have studied dimension 6 proton decay in a particular orbifold
model, where the flavour eigenstates are placed at different fixed
points.
We have found two very interesting results. \linebreak
First the predicted decay rate is enhanced compared to 4D, 
even if still compatible with the experimental bounds. In fact, for 
a cut-off scale $M_* = 10^{17} $ GeV, we set a strong lower bound on 
the compactification scale $M_c \geq 0.89\times 10^{16}\,\mbox{GeV}$,
which means that the range of validity of our model is more 
restricted also than that of 5D orbifold models~\cite{alciati}.
Secondly, the peculiar flavour structure can give striking signatures 
in the branching ratios for proton decay, suppressing strongly the 
decay into $\mu^+ K^0 $. This is due to the localization of states
in the extra-dimension and the consequent non-universal coupling of
the GUT bosons to the fermions.  

\section*{Acknowledgments}

It is a pleasure to thank my collaborators Takehiko Asaka, 
Wilfried Buchm\"uller, David Emmanuel-Costa and especially 
S\"oren Wiesenfeldt for discussions and for the fruitful and 
enjoyable collaboration.
I would also like to thank the organizers of Moriond EW '05 for
the exciting atmosphere at the workshop and for partial local
support.

\end{document}